\documentclass[12pt]{iopart}

\usepackage{iopams}
\usepackage{graphicx}
\usepackage{color}
\usepackage[hidelinks]{hyperref}

\eqnobysec

\hypersetup{colorlinks=true, linkcolor=black, citecolor=black, urlcolor=blue}

\begin{document}

\title{Cavity-enhanced magnetometer with a spinor Bose-Einstein condensate}

\author{Karol Gietka$^1$, Farokh Mivehvar$^2$,
Thomas Busch$^1$}
\address{$^1$ Quantum Systems Unit, Okinawa Institute of Science and Technology Graduate University, Onna, Okinawa 904-0495, Japan}
\address{$^2$ Institut f{\"u}r Theoretische Physik, Universit{\"a}t Innsbruck, A-6020 Innsbruck, Austria} \ead{karol.gietka@oist.jp}


\begin{abstract}
We propose a novel type of composite light-matter magnetometer based on a transversely driven multi-component Bose-Einstein condensate coupled to two distinct electromagnetic modes of a linear cavity. Above the critical pump strength, the change of the population imbalance of the condensate caused by an external magnetic field entails the change of relative photon number of the two cavity modes. Monitoring the cavity output fields thus allows for nondestructive measurement of the magnetic field in real time and we show that the sensitivity of the proposed magnetometer exhibits Heisenberg-like scaling with respect to the atom number. For state-of-the-art experimental parameters, we calculate the lower bound on the sensitivity of such a system to be of the order of  fT$/\sqrt{\mathrm{Hz}}$--pT$/\sqrt{\mathrm{Hz}}$ for a condensate of $10^4$ atoms with coherence times on the order of several ms.
\end{abstract}

\maketitle


\section{Introduction}
Being able to measure the direction, strength, and temporospatial dependence of magnetic fields with high accuracy has applications in various scientific fields, ranging from physics~\cite{edelstein2007advances,budker2007optical} and geology~\cite{tauxe2006paleomagnetic} to biology and medicine~\cite{RevModPhys.65.413,rodriguez1999perception}. High-precision magnetometers can be used to test fundamental physical theories~\cite{harry2000two,  PhysRevA.73.022107, PhysRevLett.107.171604, PhysRevLett.110.021803} and explore the boundaries of quantum metrology~\cite{RevModPhys.90.035005,PhysRevLett.96.010401,giovannetti2011advances,PhysRevLett.125.020501,doi:10.1116/5.0007577}. Until very recently, the progress in magnetometry was driven by superconducting quantum interference devices, which are based on superconducting loops containing Josephson junctions~\cite{weinstock2012squid}. However, the technological developments in cooling, trapping, and  manipulating ultracold atoms have led to the development of a next generation of atomic magnetometers~\cite{PhysRevLett.89.130801, kominis2003subfemtotesla, PhysRevLett.98.200801, PhysRevLett.104.093602, PhysRevLett.104.133601, PhysRevLett.109.253605,PhysRevLett.113.103004, PhysRevLett.111.143001, PhysRevLett.110.160802, PhysRevApplied.12.011004, PhysRevLett.124.170401, PhysRevLett.124.223001}. Among them, Faraday-rotation magnetometers~\cite{kimball2013general} have attracted a great deal of interest, as due to the possibility of increased noise suppression by creating non-classical states of light and atoms, they can operate on the verge or even beyond the standard quantum limit~\cite{PhysRevLett.111.120401,PhysRevX.5.031010}. Using anti-relaxation coatings~\cite{PhysRevLett.105.070801,PhysRevLett.104.133601}, optical multi-pass cells~\cite{PhysRevLett.110.160802}, and, finally, spin-exchange relaxation-free protocols~\cite{PhysRevLett.89.130801,kominis2003subfemtotesla} has enabled to reach sensitivities as low as 160 aT$/\sqrt{\mathrm{Hz}}$~\cite{dang2010ultrahigh}. Other state-of-the-art magnetometers rely on magnetostrictive optomechanical cavities, which can reach a peak magnetic field sensitivity of 400 nT$/\sqrt{\mathrm{Hz}}$~\cite{PhysRevLett.108.120801}, and nitrogen-vacancy centers in diamonds~\cite{maze2008nanoscale,taylor2008high,pham2011magnetic,PhysRevLett.106.080802,PhysRevLett.112.160802,PhysRevX.5.041001,barry2016optical,PhysRevApplied.8.044019}, which can exhibit sensitivities up to the order of fT$/\sqrt{\mathrm{Hz}}$.

\begin{figure}[b!]
  \centering
\includegraphics[width=\linewidth]{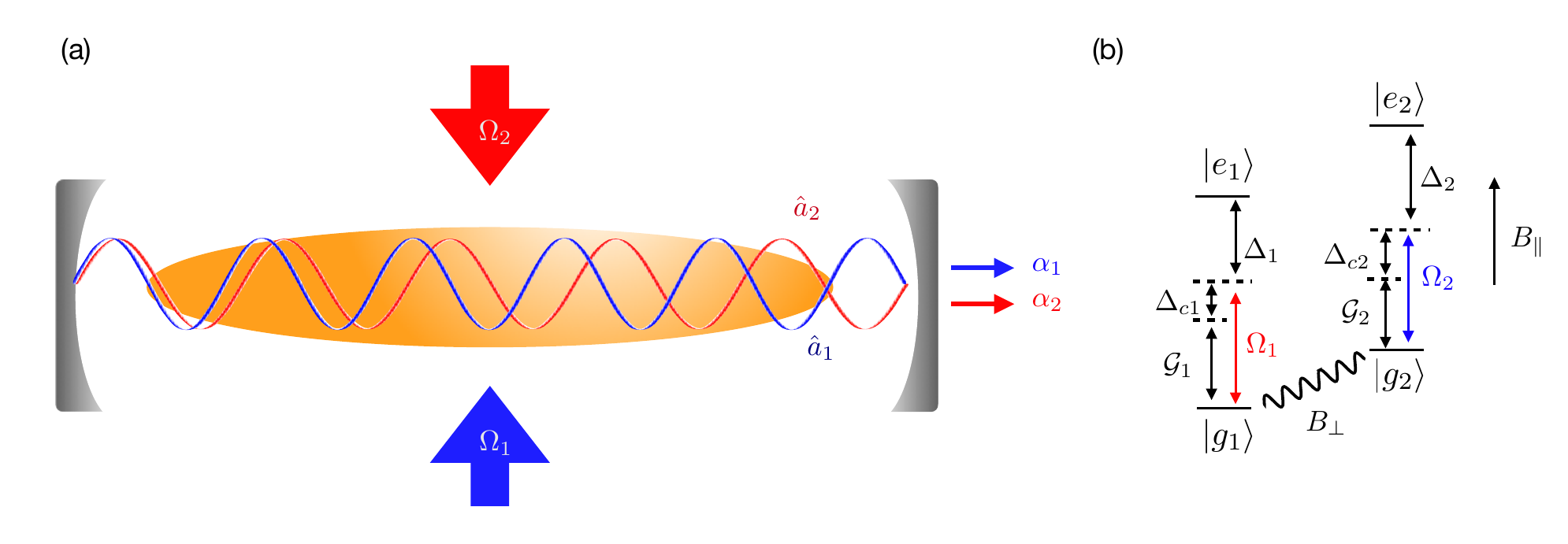}
\caption[scheme]{Schematic of the system. The atoms are strongly confined along the axis of an optical cavity and driven by two off-resonant transverse lasers with Rabi frequencies $\Omega_j$ inducing internal atomic transitions $|g_j\rangle \leftrightarrow |e_j\rangle$. These transitions are also strongly coupled to two distinct standing-wave cavity modes $\hat a_j$ with coupling strength $\mathcal G_j$. The atoms can experience a Zeeman shift due to a static magnetic field $B_\parallel$ and/or a time-dependent magnetic field $B_\perp\cos(\omega t)$, with the latter inducing transitions $|g_1\rangle \leftrightarrow |g_2\rangle$.
}
\label{fig:scheme}
\end{figure}

In this work, we propose an alternative approach to measure magnetic field strengths based on light-matter interactions in an optical cavity~\cite{RevModPhys.82.1041,RevModPhys.85.553,Mivehvar2021Cavity}. In contrast to free space, where the back-action of the particles onto the trapping laser light is negligible, in optical cavities the optical dipole force on the atoms together with the atomic back-action onto the light field give rise to complex nonlinear coupled dynamics. Motivated by recent progress in strongly coupling ultracold atoms to high-Q optical cavities~\cite{brennecke2007cavity,baumann2010dicke,Lonard1415, PhysRevLett.120.223602,PhysRevLett.121.163601,Morales2019Two, dogra2019dissipation, PhysRevLett.123.160404,kroeger2020continuous, muniz2020exploring}, we propose to harness these light-matter interactions as a sensitive probe for magnetic fields~\cite{Fan2014Hidden, Moodie2018Generalized}. The considered setup consists of a one-dimensional spinor (i.e.,~two-component) Bose-Einstein condensate (BEC) in an external (static or time-dependent) magnetic field, transversely driven by two pump lasers and dispersively coupled to two distinct electromagnetic modes of a linear cavity as depicted in Fig.~\ref{fig:scheme}~\cite{Mivehvar2017Disorder, Natalia2020Spin}. The relative occupation of the cavity modes depends on the state of the condensate, which in turn is affected by the presence of the external magnetic field. Measuring the cavity-output fields, which can be done non-destructively, then allows for real-time and continuous measurements of the magnetic field. We find that the lower bound on the sensitivity of measuring the magnetic field is on the order of fT$/\sqrt{\mathrm{Hz}}$--pT$/\sqrt{\mathrm{Hz}}$ for typical state-of-the-art experimental parameters. Although we consider this physical setting in the context of magnetometry, it can be easily extended to measurements of any field or force that can couple the spinor components of the BEC, as well as real-time monitoring of the population imbalance of a spinor condensate.


\section{Model}

The condensate we consider consists of $N$ four-level atoms with mass $M$ trapped along the axis of a two-mode standing-wave linear cavity by a tightly confining potential along the transverse directions. The atoms have two hyperfine ground states which are coherently driven from the transverse direction by two off-resonant external pump lasers, as depicted in Fig.~\ref{fig:scheme}, which induce transitions $|g_j\rangle \leftrightarrow |e_j\rangle$ ($j = 1, 2$) with the Rabi frequency $\Omega_j$. The transition $|g_j\rangle \leftrightarrow |e_j\rangle$ is also coupled to a cavity mode $\hat a_j$ with the mode function $\cos(k_{c_j} x)$ and coupling strength $\mathcal{G}_j(x) = \mathcal G_j\cos(k_{c_j} x)$, where $\mathcal G_j $ is the maximum single atom-photon coupling rate. The pump and cavity frequencies, respectively, $\omega_{p_j}$ and $\omega_{c_j} = c k_{c_j} = 2\pi c/\lambda_{c_j}$ are assumed to be near resonant with each other, but far-red detuned with respect to the atomic frequencies $\omega_{j}$. The relative energy between the two lowest lying states can be changed with a static magnetic field $B_\parallel$ that induces the Zeeman energy difference $\hbar \gamma B_\parallel$, where $\gamma$ is the gyromagnetic ratio. These states can also be coupled with an ac magnetic field $B_\perp \cos \omega t$, introducing an (ac Zeeman) energy shift $\hbar\omega$, and inducing magnetic dipole transition between $|g_1\rangle$ and $|g_2\rangle$ with a magnetic Raman frequency $\Omega = \gamma B_\perp$ .

In the dispersive regime $|\Delta_j| \equiv |\omega_{p_j} - \omega_j| \gg \{\Omega_j; \mathcal{G}_j\}$, in which the atomic excited states $|e_j\rangle$ quickly reach a steady state, their dynamics can be adiabatically eliminated (see \ref{sec:eff-model}). As a result, in the rotating frame of the pumping lasers, the effective Hamiltonian for the atoms in the ground states and the cavity fields becomes 
\begin{equation} \label{eq:H}
\hat{H}=
\int \hat\Psi^\dag(x)\hat{\mathcal{H}}_0\hat\Psi(x)\,\mathrm{d}x-\hbar\Delta_{c_1}\hat{a}_1^\dagger\hat{a}_1 -\hbar\Delta_{c_2}\hat{a}_2^\dagger\hat{a}_2,
\end{equation}
where $\hat \Psi = (\hat \psi_1, \hat\psi_2)^\intercal$ is the bosonic spinor field annihilation operator and $\hat{\mathcal{H}}_0$ is the single-particle Hamiltonian density
\begin{equation} \label{eq:H0}
 \hat{\mathcal{H}}_0 =
    \left( \matrix{\frac{\hat{p}^2}{2M}+\hat{V}_1(x)  -\hbar\gamma B_\parallel -\hbar \omega & i \hbar \Omega/2
     \cr
     - i \hbar \Omega/2 & \frac{\hat p^2}{2M}+\hat{V}_2(x) +\hbar \delta}\right).
\end{equation}
Here, $\hat{V}_{j}(x) = \hbar U_{j}\hat a_j^\dagger \hat a_j\cos^2(k_{c_j} x) + \hbar \eta_j(\hat a_j +\hat a_j^\dagger)\cos(k_{c_j} x)$ is the dynamical cavity potential with $\hbar U_{j} = \hbar \mathcal{G}_j^2/\Delta_j$ being the maximum depth of the optical potential per photon due to the absorption and emission of cavity photons. The maximum depth of the optical potentials due to the redistribution of photons between the pump lasers and the cavity fields are given by $\hbar \eta_j = \hbar \mathcal{G}_j \Omega_j /\Delta_j$, and $\delta = \Omega^2_2/\Delta_2 - \Omega^2_1/\Delta_1 +\omega_{12}$ is the Stark-shifted two-photon detuning, with $\omega_{12}$ being the bare energy difference between $|g_1\rangle$ and $|g_2\rangle$. For the sake of clarity, in remainder of this work we will focus on the balanced condition, {i.e.}, $U_0 \equiv U_1 = U_2 $, $\Delta_c \equiv \Delta_{c_1}=\Delta_{c_2}$, $\omega_{c}\equiv \omega_{c_1}=\omega_{c_2}$, $k_{c}\equiv k_{c_1}=k_{c_2}$, and $\eta_0\equiv \eta_1 = \eta_2 $. We assume that the atom-atom interactions are negligible with respect to the cavity-mediated interactions, which is quantitatively a good approximation for spinor-BEC--cavity-QED experiments~\cite{PhysRevLett.123.160404,PhysRevLett.121.163601}.

In the mean-field approximation, the system is the described by a set of four coupled equations for the cavity-field amplitudes $\langle \hat a_j(t) \rangle = \alpha_j (t)$ and the atomic condensate wave functions $\langle \psi_j(x,t) \rangle = \psi_j(x,t)$, given by
\begin{eqnarray}\label{eq:eoms1}
  i \frac{\partial}{\partial t} \alpha_j = \left[-\Delta_c + U_0 \langle \cos^2(k_c x)\rangle_j- i \kappa\right]\alpha_j + \eta_0\langle \cos(k_c x) \rangle_j,\nonumber \\
  i \hbar \frac{\partial}{\partial t} \psi_1 =\left[\frac{p^2}{2M}+V_1(x)-\hbar(\gamma B_\parallel + \omega) \right]\psi_1 +i\frac{\Omega}{2} \psi_2,\nonumber \\
  i \hbar \frac{\partial}{\partial t} \psi_2 =\left[\frac{p^2}{2M}+V_2(x)+\hbar \delta\right]\psi_2 -i\frac{\Omega}{2} \psi_1,
\end{eqnarray}
where $V_j(x)= \langle \hat{V}_j(x) \rangle$, and $\langle \cos^2(k_c x)\rangle_j$ and $\langle \cos(k_c x)\rangle_j$ are spatial overlaps of $\cos^2(k_c x)$ and $\cos(k_c x)$ with the corresponding atomic density $|\psi_j(x)|^2$, respectively. Here we have introduced the cavity-photon loss rate $\kappa$ which signifies the open nature of the system and is a crucial component in the model~\cite{dogra2019dissipation}. It provides a way for the system to reach a steady state and allows for non-destructive monitoring of this state. 

It can be shown that this system can also be described by two coupled Dicke models~\cite{PhysRevLett.104.130401}, whose two low-lying polaritons are coupled by an external field \cite{PhysRevLett.95.010402,abbarchi2013macroscopic}. Since the effective spin formed by these two polaritons is not coupled to the cavity field (no optical $|g_1\rangle \leftrightarrow |g_2 \rangle$ transition), the only mechanism leading to the relaxation of the effective spin will be spontaneous emission, allowing for long coherence times and consequently long measurement times \cite{PhysRevLett.118.063604}.

\begin{figure*}[tb!]
  \centering
\includegraphics[width=0.493\textwidth]{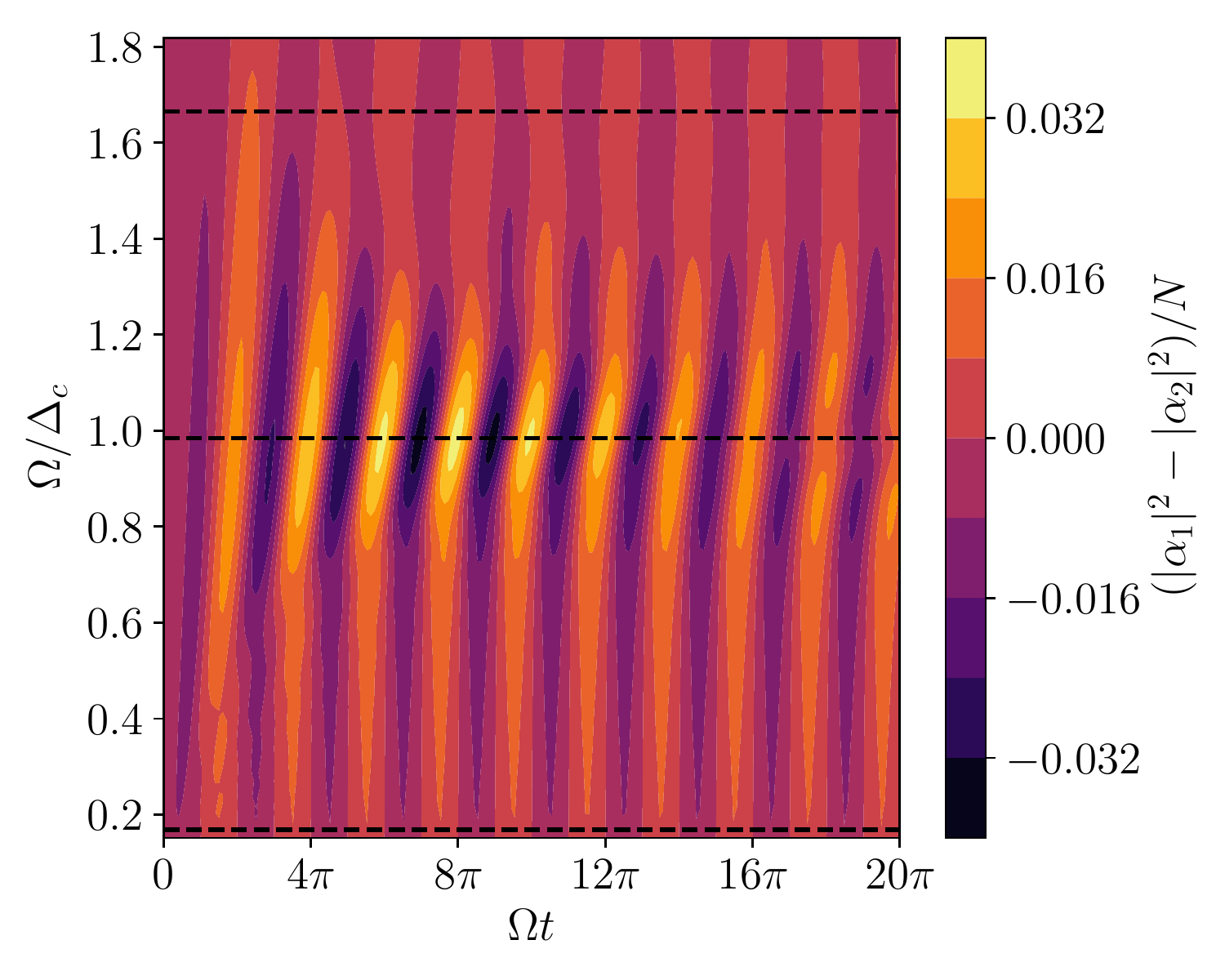}
\includegraphics[width=0.493\textwidth]{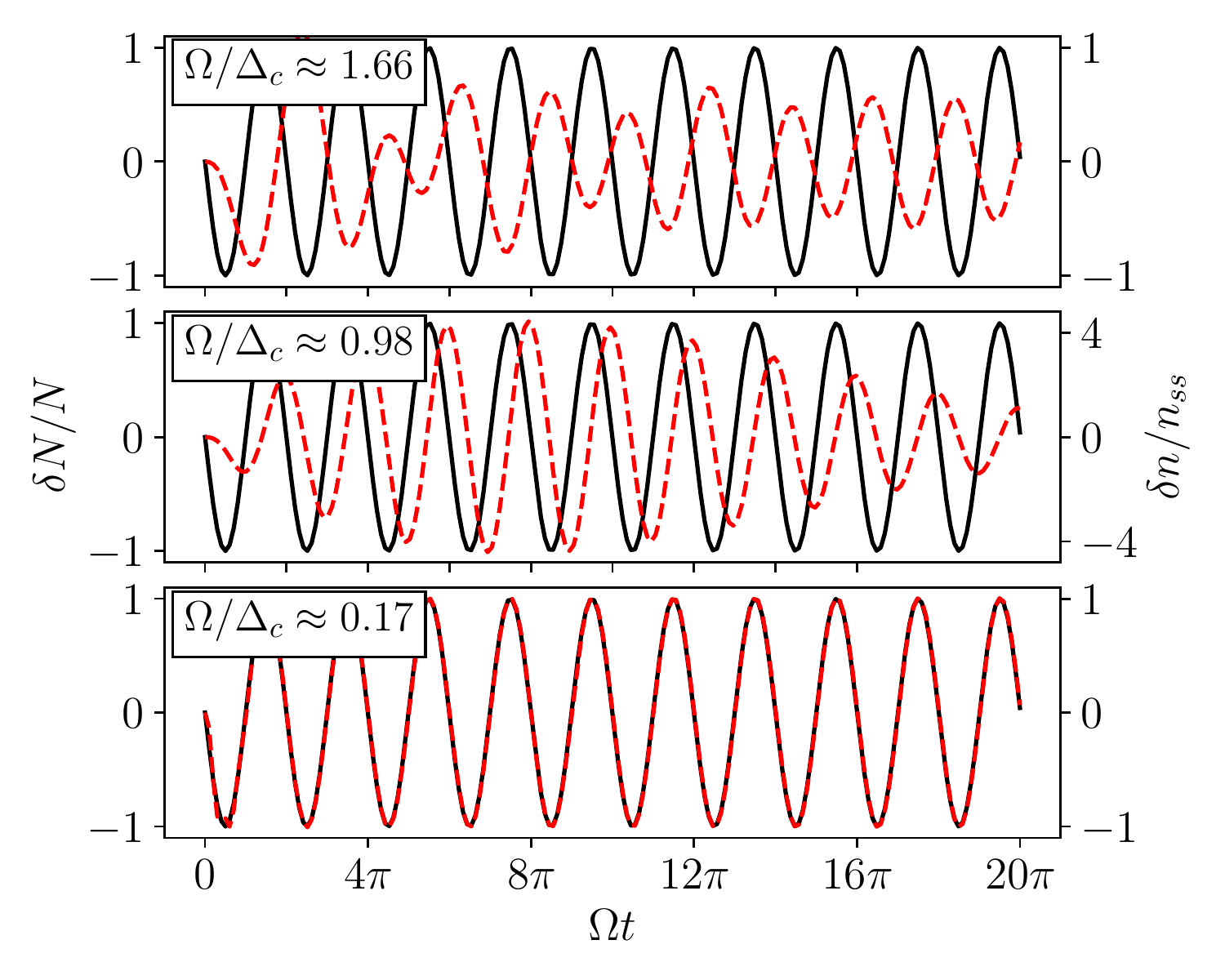}
\caption[photon_oscillation]{Left hand side: The normalized relative photon number $(|\alpha_1|^2-|\alpha_2|^2)/N$ as a function of the magnetic Rabi frequency $\Omega$ scaled by the cavity detuning $\Delta_c$ and time $t$ normalized to the Rabi frequency $\Omega$. An increasing shift between the oscillations of the spinor BEC and the relative average number of photons can be clearly seen: depending on the relative strength of $\Omega$ and $\Delta_c$, one can identify three distinct photon-dynamics regions for $\Omega < |\Delta_c|$, $\Omega \sim |\Delta_c|$, and $\Omega > |\Delta_c|$ (indicated by horizontal-black-dashed lines). Examples of these are presented on the right-hand side, where the solid-black line represents the normalized BEC population-imbalance oscillations and the dashed-red line represents the relative photon number oscillations normalized to the maximal number of scattered photons in the steady state $n_{ss}$ [see equation~(\ref{eq:ssnop})]. The dashed-black lines on the left panel indicate the cuts presented on the right-hand side. See the text for detailed explanation. The parameters are set to $(\Delta_c,U_0,\eta_0,\kappa) = (-3300,-1/600,300,300)\omega_r$ and $N = 10^4$.}
\label{fig:photonosc}
\end{figure*}

\section{Magnetometry}

We first consider a Ramsey scheme, for which we have $\{\Omega,\omega\} = 0$ \cite{PhysRev.78.695}. The condensate is initially prepared in an equal superposition state ($\delta N \equiv N_1-N_2 = 0$ with $N_j = \int |\psi_j(x,t)|^2\, \mathrm{d}x$), and we subsequently turn on the static magnetic field $B_\parallel$ which induces a Zeeman splitting $ \hbar \gamma B_\parallel$ of the energy levels. In this static magnetic field the two spinor components acquire a relative phase $\phi = \tau \gamma B_\parallel$\footnote{For clarity, we have moved to the reference frame rotating with frequency $\delta$, so the accumulated phase is not $\phi = \tau(\gamma B_\parallel +\delta)$, but $\phi = \tau\gamma B_\parallel$.}, where $\tau$ is the interrogation time. After switching off the static magnetic field, a $\pi/2$ pulse is applied which converts the relative phase into a relative atom number. This can be easily seen if we introduce total pseudo-spin operators defined as $\hat{\mathbf{s}} = \int \hat{\Psi}^\dagger \vec{\sigma} \hat{\Psi}\,\mathrm{d}r$, where $\vec{\sigma}$ is the vector of Pauli matrices. In particular, $\hat s_z = N_1-N_2=\delta N$ corresponds to the population imbalance. Any unitary transformation of such a pseudo-spin can be depicted as a rotation $\exp(- i \phi \hat s_{\mathbf{n}}/2 )$ on the generalized Bloch sphere, where $\mathbf{n}$ and $\phi$ are the rotation axis and rotation angle, respectively. Introducing a pseudo-spin state $|\vartheta, \varphi \rangle \equiv \sqrt{N}\left(e^{i \varphi}\cos(\vartheta/2)|g_1 \rangle + \sin(\vartheta/2)| g_2 \rangle\right)= \int \hat \Psi\,\mathrm{d}x$, where $\vartheta$ and $\varphi$ are the azimuthal and polar angles, the conversion of the relative phase to the relative atom number ($\langle \vartheta,\varphi | \hat s_z| \vartheta,\varphi \rangle = N \cos \vartheta$) can be conveniently expressed by $ e^{- i \hat s_x \pi/4} |{\pi}/2,\phi \rangle = |{\pi}/{2} -\phi, 0 \rangle$. We note that in the superradiant regime where the strength of the effective cavity pump $\eta_0$ is above a certain threshold, and which is the regime we focus on here, the average number of photons $|\alpha_j|^2$ in cavity mode $j$ is proportional to the atom number $N_j$ in pseudo-spin state $j$; cf.\ equation~(\ref{eq:eoms1}). By measuring the relative photon number $|\alpha_1(\phi)|^2 - |\alpha_2(\phi)|^2 \equiv \delta n$ during or after a time interval $t$, one can then estimate the relative atom number $\delta N$ and thus $\phi$ and the strength of the static magnetic field.

To measure an oscillating magnetic field $B_\perp \cos \omega t$, one can employ a Rabi scheme~\cite{PhysRev.51.652}. In the rotating frame of the oscillating magnetic field and in the presence of a small bias field, $B_\parallel$ (fixing the axis of quantization), the relative energy between the two ground states is shifted by an amount $\hbar \omega$. The oscillating field induces Rabi oscillations between the two components of the BEC with a frequency $\Omega = \gamma B_\perp$ [see equation~(\ref{eq:H0})], and for the resonant case $ -\gamma B_\parallel - \omega = \delta$, the spinor components will oscillate without acquiring any relative phase. In this case the entire information about the amplitude of the magnetic field $B_\perp$ is encoded in the period of the spinor oscillations, i.e., the oscillations of the relative number of atoms. Since the photon scattering probability depends on the number of atoms, the spinor oscillations directly lead to oscillations of the cavity mode amplitudes with frequency $\Omega$ as well. In Fig.~\ref{fig:photonosc} we show the normalized relative photon number $(|\alpha_1|^2-|\alpha_2|^2)/N$ as a function of (unitless) time $\Omega t$ and (unitless) frequency $\Omega/\Delta_c$. In this graph one can clearly distinguish three distinct regimes of photon scattering. When $\Omega < |\Delta_c|$, the optical potential adiabatically follows the atomic density oscillations. For $\Omega \sim |\Delta_c|$, the photon dynamics cannot completely follow the BEC population-imbalance oscillations, introducing a time delay between the BEC population-imbalance and cavity field-amplitude oscillations. Surprisingly, the number of scattered photons increases in this regime, leading to enhanced sensitivity with respect to the adiabatic case. Finally, when $\Omega > |\Delta_c| $, the photon scattering cannot keep up with the Rabi oscillations of the condensate, introducing thus a constant $\pi$ phase shift between atomic and photonic oscillations. Hence, the number of scattered photons decreases until the relative atom number oscillations are so fast that they no longer affect scattering of photons (see \ref{sec:cavity-field-ss} for the details).


\section{Measurement back-action}

\begin{figure}[tb!]
  \centering
\includegraphics[width=0.493\textwidth]{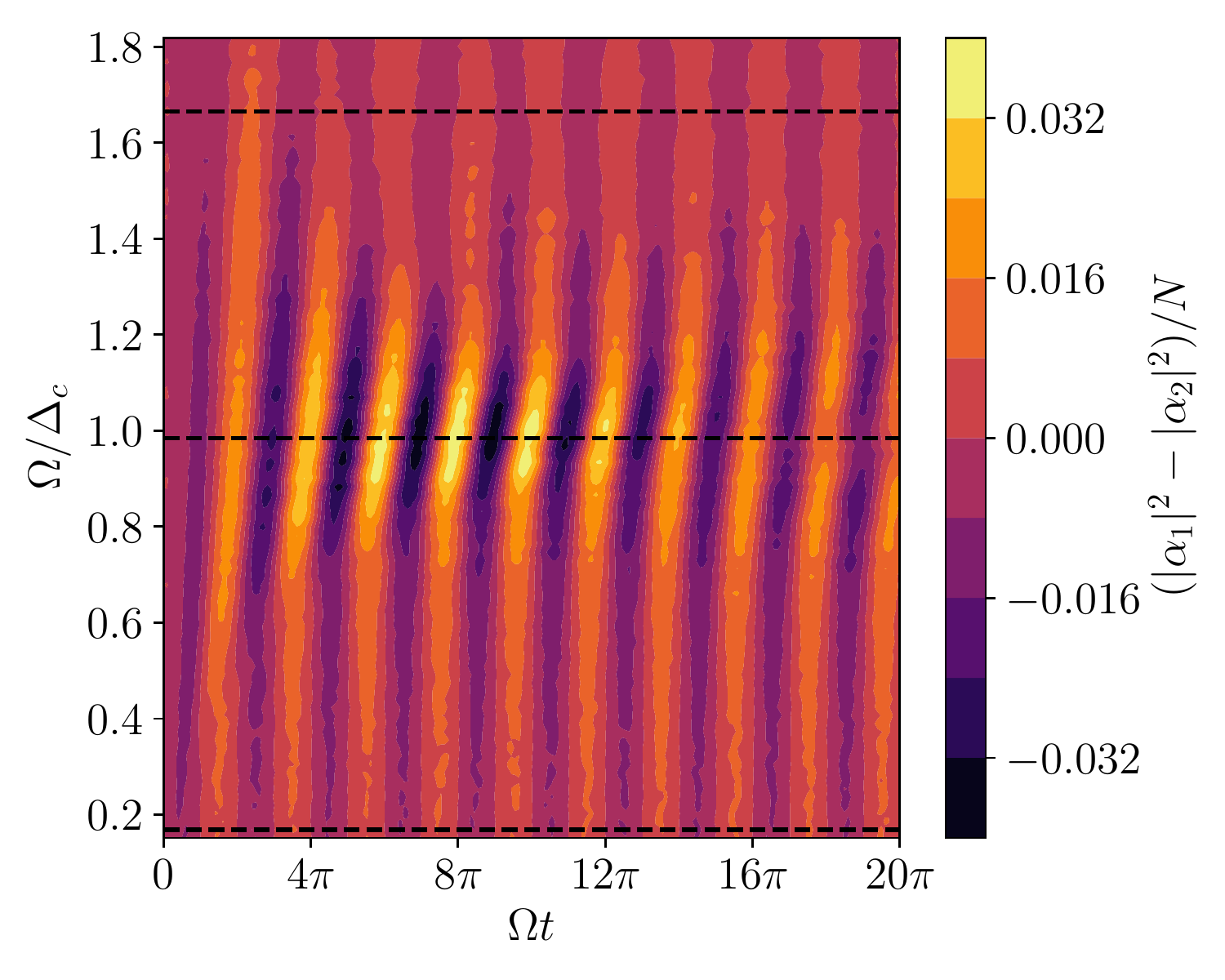}
\includegraphics[width=0.493\textwidth]{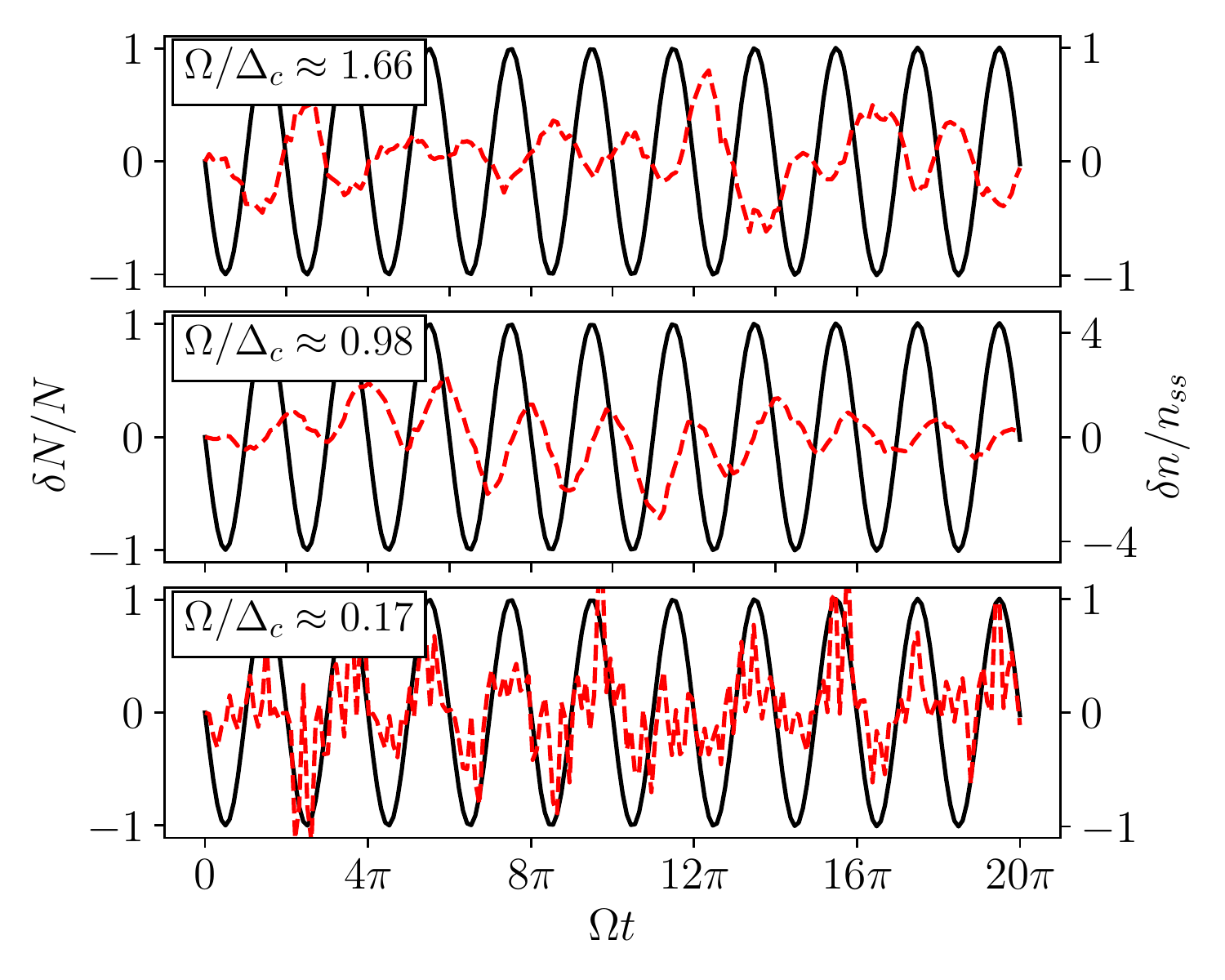}
\caption[photonoscillationstoch]{ The effect of continuous measurement back-action on the cavity-field dynamics. Here we set the measurement efficiency equal to 50\% ($\epsilon = 0.5$). Although the measurement back-action strongly affects single quantum trajectories (right), by averaging over many independent trajectories (for the calculations we used 100 trajectories) the noise averages out (left); cf.\ Fig.~\ref{fig:photonosc}. The parameters for this simulation are the same as in Fig.~\ref{fig:photonosc}.}
\label{fig:photonoscstoch}
\end{figure}

Even though our system allows for non-destructive measurement, an inevitable consequence of any measurement is the measurement back-action. This generally adds noise to the system and thus to the measurement outcomes themselves, which reduces the estimation precision \cite{RevModPhys.52.341}. In general, the measurement-back action is a stochastic process and, in the limit of continuous measurement, the evolution of the system state $\hat \rho$ is governed by the stochastic master equation \cite{jacobs2006straightforward}
\begin{equation}
\mathrm{d}\hat{\rho} = - \frac{i}{\hbar} \left[ \hat H, \hat{\rho} \right]\mathrm{d}t +\mathcal{D}[\hat{c}]\hat{\rho}\,\mathrm{d}t + \sqrt{\epsilon}\mathcal{H}[\hat{c}]\hat{\rho}\,\mathrm{d}W.
\end{equation}
Here we have defined the Lindblad superoperator
\begin{equation}
\mathcal{D}[\hat{c}] : = \hat c \hat{\rho}\hat c^\dagger-\frac{1}{2}\left(\hat c^\dagger \hat c \hat{\rho} + \hat{\rho}\hat c^\dagger \hat c \right),
\end{equation}
 the measurement superoperator
\begin{equation}
\mathcal{H}[\hat{c}] : = \hat c \hat{\rho} + \hat{\rho}\hat c^\dagger -\langle \hat c + \hat{c}^\dagger \rangle \hat{\rho},
\end{equation}
the Wiener process increment $\mathrm{d}W$, and a measurement efficiency $\epsilon \leq 1$.
The Lindblad superoperator describes the disturbance of the system due to the measurement, and the measurement superoperator represents the gain of information resulting from a measurement. For an arbitrary operator $\hat O$, we can combine the stochastic master equation and $\mathrm{d}\langle \hat O \rangle  = \mathrm{Tr}[\hat O, \mathrm{d} \hat{\rho}]$ to obtain the equation of motion for the expectation value of $\hat O$ as
\begin{eqnarray}
  \mathrm{d}\langle \hat O \rangle  =  - \frac{i}{\hbar}\left\langle\left[\hat O, \hat H \right]\right\rangle\mathrm{d}t&+ \left\langle \hat c^\dagger \hat O \hat c -\frac{1}{2}\left(\hat c^\dagger \hat c \hat O + \hat O \hat c^\dagger \hat c\right)\right\rangle \mathrm{d}t\nonumber 
\\ 
&+ \sqrt{\epsilon}\left\langle \hat c^\dagger \hat O  +\hat O \hat c -\left \langle \hat O \right\rangle \left\langle\hat c^\dagger + \hat c \right \rangle \right\rangle \mathrm{d}W.
\end{eqnarray}
For a homodyne detection scheme, which can be used to measure the average number of photons \cite{PhysRevA.73.033808}, we have $\hat c = \sqrt{2\kappa} \hat a$, and if we set the measurement efficiency equal to zero, it is straightforward to re-derive equation~(\ref{eq:eoms1}). Since the atomic field operators $\hat \psi_j$ commute with the cavity field operators $\hat a_j$, a non-zero measurement efficiency modifies the mean-field equations for the cavity fields as
  \begin{equation} 
  i \frac{\partial}{\partial t} \alpha_j = \left[-\Delta_c + U_0 \langle \cos^2(k_c x)\rangle_j\right]\alpha_j + \eta_0\langle \cos(k_c x) \rangle_j- i \kappa \alpha_j +i \sqrt{2 \epsilon \kappa} \xi(t),
\end{equation}
where $\xi(t)$ is white noise. In other words, in the mean-field limit the measurement back-action adds a stochastic noise to the cavity fields. This in turn leads to the deformation of the optical potentials inside the cavity and eventually to heating and loss of atoms. In the case of the Ramsey scheme, this can be counteracted by a feedback loop \cite{kroeger2020continuous} or included in the process of calibration. In the case of the Rabi scheme, the effect of the measurement back-action should be more pronounced as the number of scattered photons is a function of time. In order to obtain the average relative photon number, one would therefore have to repeat the experiment many times. Because the fluctuations of the optical potentials do not change the relative number of atoms in the condensate, our scheme realizes a non-demolition quantum measurement of condensate state. This can be seen in Fig.~\ref{fig:photonoscstoch}, where we have plotted the  averaged and single quantum trajectories for different noise realizations of the relative average number of photons as a function of $\Omega$ for $\epsilon = 0.5$. 

\section{Cavity-field measurement and sensitivity}
After carrying out the full Ramsey scheme and reaching a stationary state, the relative average number of photons $\delta n$ is measured for a time interval $t$ and the value of the magnetic field $B_\parallel$ is estimated from the collected data gathered by performing enough measurements to average out the effect of noise. The relative average number of detected photons during $t$ will be
\begin{equation}
  \delta n(\phi) = \kappa t |\alpha_0|^2 \cos \phi,
\end{equation}
where $|\alpha_0|^2$ is the maximum number of scattered photons in a single mode. Subsequently, the value of $\phi$ is estimated with the sensitivity given by the error propagation formula
\begin{equation} \label{eq:errorprop}
  \Delta \phi = \frac{\Delta\left[\delta n (\phi)\right]}{\partial_\phi \left[\delta n (\phi)\right]} = \frac{\Delta |\alpha_1 (\phi)|^2+ \Delta |\alpha_2 (\phi)|^2}{\kappa t |\alpha_0|^2 |\sin \phi |} .
\end{equation}
For classical fields, the uncertainty of the relative average number of photons can easily be estimated as $\Delta \left[\delta n (\phi)\right] \approx \sqrt{\kappa t}|\alpha_0|$, and the sensitivity for measuring $B_\parallel$ becomes
\begin{equation}\label{eq:Ramsey}
  \Delta B_\parallel =\frac{1}{\gamma \tau\sqrt{\kappa t} |\alpha_0 \sin \phi |} \geq \frac{1}{\gamma \tau\sqrt{\kappa t}|\alpha_0|}.
\end{equation} 

In the Rabi scheme case, even though the qualitative behavior of cavity fields is different in each regime of parameters (see \ref{sec:cavity-field-ss}), the relative average number of photons inside the cavity oscillates with frequency $\Omega$ as illustrated in Fig.~\ref{fig:photonosc}. Therefore, measuring the cavity output fields should allow for nondestructive measurement of $\Omega$. We assume that the cavity field is detected over short time intervals $\delta t$ during which the cavity output field is approximately constant and the measurement is repeated enough times to smooth out the fluctuations of the field. Then the {\it{instantaneous}} relative average number of detected photons at time $t$ is
\begin{equation}\label{eq:rabphot}
  \delta n(t)  \approx \kappa \int_{t-\delta t/2}^{t+\delta t/2}|\alpha_0|^2\cos \Omega t^\prime\, \mathrm{d}t^\prime  
  \approx \kappa \delta t |\alpha_0|^2 \sin \Omega t. 
\end{equation}
Again, by comparing the measured data with the model in equation~(\ref{eq:rabphot}), the value of the magnetic field $B_\perp$ can be estimated with the uncertainty given by the error propagation formula
\begin{equation}\label{eq:Rabi}
  \Delta B_\perp \approx \frac{1}{\gamma t \sqrt{\kappa\delta t}|\alpha_0\sin(\Omega t)|} \geq \frac{1}{\gamma t \sqrt{\kappa\delta t}|\alpha_0|}.
\end{equation}

It is also instructive to refer to the quantum theory of estimation and calculate the quantum Fisher information, $F_q$, which sets a lower bound on the sensitivity through the quantum Cram{\'e}r-Rao bound as $\Delta B \geq 1/\sqrt{F_q}$~\cite{PhysRevLett.72.3439, PhysRevLett.102.100401}. For pure states the Fisher information is defined as $F_q = 4 (\Delta \hat h)$ with $ \hat h=i[\partial_{B} \hat{\mathcal{U}]}\hat{\mathcal{U}}^\dagger$ being the generator of an infinitesimal change along a trajectory parametrized by $B$. From the viewpoint of the relative average number of cavity photons $\hat a = (\hat a_1 - \hat a_2)/2$ and in the reference frame in which $\alpha_1,\alpha_2 \in \Re$\footnote{We move to such a frame of reference for the clarity of calculations.}, the dynamics will be governed by the displacement operator
\begin{equation}
  \hat{\mathcal{U}}(t) = \hat D(\alpha) = \exp\left(\alpha \hat a^\dagger - \alpha^* \hat a \right),
\end{equation}
where $\alpha = \alpha^*\equiv \alpha(t)$ depends on the regime of photon scattering. The generator of the infinitesimal change $ \hat h$ can be easily calculated as
\begin{equation}
  \hat h = i \frac{\partial \alpha}{\partial B}\left(\hat a^\dagger - \hat a \right),
\end{equation}
and from this the quantum Fisher information can be directly determined to be given by
\begin{equation}\label{eq:qfi}
  F_q = 4 (\Delta \hat h)^2 = \left( \frac{\partial \alpha}{\partial B} \right)^2.
\end{equation}
It is easy to show that for both, the Ramsey scheme [$\alpha\approx |\alpha_0|\sqrt{\kappa t } \cos \left(\gamma B_\parallel \tau \right)$] and the Rabi scheme [$\alpha \approx |\alpha_0| \sqrt{ \kappa \delta t}\cos(\gamma B_\perp t)$], the quantum Cram{\'e}r-Rao bound yields the same limitation on the sensitivity as equations~(\ref{eq:Ramsey}) and~(\ref{eq:Rabi}).

Finally, let us make a comment concerning the average number of photons. The average number of cavity photons can be calculated using the steady-state solution. Deep in the self-ordered (superradiant) regime~\cite{nagy2008self}, we get
\begin{equation}\label{eq:ssnop}
  n_{ss}\equiv|\alpha_0|^2 \approx \frac{N^2\eta^2}{\left[\Delta_c - N U_0\right]^2+\kappa^2},
\end{equation}
where $\Delta_c -NU_0$ is the dispersively shifted cavity detuning. Therefore, for negligible dispersive shift $\Delta_c - N U_0 \approx \Delta_c $, the average number of photons will be $|\alpha_0|^2\sim N^2\eta^2/(\Delta_c^2 +\kappa^2)$, and from the viewpoint of the atoms, we obtain a Heisenberg-like scaling of the sensitivity similarly to what was found in~\cite{PhysRevLett.122.190801}. In two-mode systems with a fixed number of particles, for example a collection of $N$ interacting two-level atoms, the Heisenberg scaling means that the sensitivity scales like $\sim 1/N$. In the systems composed of solely atoms this can only be achieved with the use of entanglement. The achieved $\sim 1/N$ scaling in our proposal, however, has its origins not in entanglement but in superradiance, which is a consequence of strong light-matter interaction. The information about the magnetic field is encoded linearly into the state of the atoms which are not entangled. Performing thus an optimal measurement directly on such atoms would give rise maximally to the standard quantum limit $\sim 1/\sqrt{N}$. That said, due to the strong coupling of the atoms to the cavity modes, the information about the magnetic field is  also effectively imprinted into the state of the cavity fields. Since the average number of photons in the superradiant regime is proportional to the number of atoms squared $n_{ss} \sim N^2$, measuring the cavity fields gives rise to a $1/N$ scaling of the sensitivity. Because the origin of this scaling is not entanglement, we call it Heisenberg-like scaling.

It is worth noting that since in the non-superradiant regime no photons is scattered into the cavity modes, the presented scheme requires to operate in the superradiant phase.


\section{Implementation and estimated sensitivity}
This magnetometer proposal is based on existing experimental technologies for quantum-gas-cavity systems~\cite{PhysRevLett.120.223602,PhysRevLett.121.163601,dogra2019dissipation,PhysRevLett.123.160404,kroeger2020continuous}. For instance, the spinor BEC could be realized by coupling two internal states of $^{87}$Rb,  $| F,m_f\rangle = |1,-1\rangle \equiv |\!\downarrow\,\rangle$ and $| F,m_f\rangle = |2,-2\rangle \equiv |\!\uparrow\,\rangle$ to the cavity modes, similarly as in Refs.~\cite{PhysRevLett.121.163601,PhysRevLett.123.160404}, or by making use of the  $F=1$ total angular momentum manifold of a $^{87}$Rb by preparing the gas in the mixture of $m_f = +1$ and $m_f = -1$ sub-levels as in Ref.~\cite{PhysRevLett.120.223602}. Coupling the spins to two different cavity modes could be realized by using two orthogonal polarizations~\cite{Morales2019Two} or two modes with different frequencies~\cite{faghihi2013entanglement}.
We, therefore, attempt in the following to estimate the sensitivity lower bound by using state-of-the-art experimental parameters 
from recent spinor-BEC-cavity experiments~\cite{PhysRevLett.120.223602,dogra2019dissipation}. Assuming that one performs $m = T/t_s$ experiments with $T$ being the total time of the experiment and $t_s$ being the time of a single experiment which may consist of many measurements, for Ramsey scheme we obtain
\begin{equation}\label{eq:ramseysens}
  \Delta B_\parallel \sqrt{T} =\frac{\sqrt{t_s}}{\gamma \tau \sqrt{\kappa t}|\alpha_0|},
\end{equation}
and for Rabi scheme we find
\begin{equation}\label{eq:rabisens}
  \Delta B_\perp \sqrt{T} =\frac{\sqrt{t_s}}{\gamma t \sqrt{\kappa \delta t}|\alpha_0|}.
\end{equation}
Assuming $^{87}$Rb atoms with a gyromagnetic ratio of $\gamma = 2\pi \times 7$ $\mathrm{Hz}/\mathrm{nT}$, an experimental cycle time of the order of 1s, a coherence time of the order of 10ms, and ideal photon detectors, the lower bound on the sensitivity of the cavity-based magnetometer containing around $ 10^4$~$^{87}$Rb atoms would be $\Delta B_\parallel \sqrt{T} \sim \mathrm{fT}/\sqrt{\mathrm{Hz}}$ for a Ramsey scheme and $\Delta B_\perp \sqrt{T} \sim 10$ $ \mathrm{ pT}/\sqrt{\mathrm{Hz}}$ for the Rabi scheme. However, under realistic conditions, {i.e.}, taking into account experimental and measurement back-action noise, limitations associated with heating the condensate by strong intra-cavity lattices \cite{kroeger2020continuous}, and dispersion of the cavity deforming the interference lattice, a realistic sensitivity will be much lower. Taking the average number of photons from~\cite{PhysRevLett.123.160404} the sensitivity of the proposed magnetometer would rather be on the order of $\sim10$ $\mathrm{pT}/\sqrt{\mathrm{Hz}}$ for the Ramsey scheme and $\sim100$ $\mathrm{nT}/\sqrt{\mathrm{Hz}} $ for the Rabi scheme, which is comparable with other state-of-the-art magnetometers. The proposed magnetometer should be also possible to be implemented in ensembles of thermal atoms which realize two coupled Dicke systems~\cite{Fan2014Hidden, Moodie2018Generalized, Morales2019Two}.

In principle, one can also loosen the balanced condition, and consider a limiting case where one measures only a single mode of the cavity, $\hat a_j$. In this case the intra-cavity field amplitude will behave as
\begin{equation}
  |\alpha_j| \approx \frac{|\alpha_0|}{2}\left[1 \pm \cos(\phi) \right].
\end{equation}
Inserting this expression into the formula for the quantum Fisher information~(\ref{eq:qfi}), it is straightforward to show that the lower bound of the sensitivity will be two times higher than the bounds from equations~(\ref{eq:Ramsey}) and (\ref{eq:Rabi}).
\newline

\section{Outlook and conclusions}
We have presented an experimentally realistic scheme for the highly-precise measurement of a magnetic field strength based on light-matter interaction inside a two-mode linear cavity. On the one hand, the back-action of the atoms onto the light fields transfers the information about the coupling to the cavity field, and on the other hand, the cavity fields help to trap the atoms in many lattice sites. Importantly this also allows to non-destructively monitor the dynamics of the system. Since the proposed magnetometer operates in the superradiant regime, it, therefore, exhibits a Heisenberg-like scaling of the sensitivity.

We have found that the lower bound on the sensitivity of such a magnetometer is on the order of $\sim\mathrm{fT}/\sqrt{\mathrm{Hz}}$ for a Ramsey scheme and $\sim10$ $\mathrm{pT}/\sqrt{\mathrm{Hz}}$ for the Rabi scheme, assuming total measurement times on the order of 1 s and condensates with about $10^4$ atoms. The concept of the proposed magnetometer is built upon the fact that a change of the relative occupation of the spinor components leads to a change of the relative average number of photons which can be measured through the cavity-output fields. Although we have shown how to exploit the light-matter interaction for magnetometry, the scope of this physical setting can be easily extended to measure any type of field or force that can modify or couple spinor components of the BEC. From a different perspective, the proposed setup allows to monitor the state and dynamics of a spinor BEC in real time. An interesting extension would be to additionally take advantage of light-matter interaction to create non-classical states of matter and/or light and further increase the sensitivity of the proposed machine~\cite{RevModPhys.90.035005,PhysRevLett.104.250801,PhysRevLett.110.120402,PhysRevLett.125.200505,gietka2017quantum}. One can easily imagine that number squeezing of the cavity fields would reduce the sensitivity below the photon shot-noise limit. Unfortunately, a full quantum treatment of such a system would require gargantuan computational power and, except for few body problems and semi-classical models, it is currently beyond the scope of theoretical investigations~\cite{feynman1999simulating}.


\ack
Simulations were performed using the open-source QuantumOptics.jl framework in Julia~\cite{kramer2018quantumoptics}. K.G. would like to acknowledge discussions with Stefan Ostermann, David Plankensteiner, Helmut Ritsch, Juan Polo Gomez, James Kwiecinski, Jan Ko{\l}ody\'nski and also correspondence with Tobias Donner and Ronen Kroeze. This work was supported by the Okinawa Institute of Science and Technology Graduate University. K.G. acknowledges support from the Japanese Society for the Promotion of Science (P19792). F.\,M.\ is supported by the
Lise-Meitner Fellowship M2438-NBL of the Austrian Science Fund (FWF), and the International Joint Project No.\ I3964-N27 of the FWF and the National Agency for Research (ANR) of France.

\setcounter{section}{1}
\appendix

\section{Effective model}
\label{sec:eff-model}

\noindent The energies of the states of the four-level bosonic atoms are $\hbar\omega_{g_1} = -\hbar\gamma B_\parallel, \hbar\omega_{g_2} , \hbar\omega_{e_1}$, and $\hbar\omega_{e_2}$ (see the main text for details). In the dipole and the rotating wave approximations, the single-particle Hamiltonian density becomes
\begin{equation}
\eqalign{ \fl
 \hat{\mathcal{H}} = \frac{\hat p^2}{2M}I_{4\times4} +\sum_{\xi =\{g_1,g_2,e_1,e_2\}} \hbar \omega_\xi \hat \sigma_{\xi \xi}+ \hbar \omega_{c_1}\hat a^\dagger_1 \hat a_1 + \hbar \omega_{c_2}\hat a^\dagger_2 \hat a_2 \\ + \hbar \Big[\mathcal{G}_1(x)\hat a_1 \hat \sigma_{e_1g_1} + \mathcal{G}_2(x)\hat a_2 \hat \sigma_{e_2g_2} + \mathrm{H.c.} \Big] \\
 + \hbar\Big[ \Omega_1 e^{-i \omega_{p_1} t}\hat \sigma_{e_1g_1} +\Omega_2 e^{-i \omega_{p_2}t}\hat\sigma_{e_2g_2} +\mathrm{H.c.}\Big] + \hbar \Big[ \Omega e^{- i \omega t }\hat \sigma_{g_1g_2} + \mathrm{H.c.}\Big],}
\end{equation}
where $m$ is the atomic mass, $\hat a_j$ is the annihilation operator of cavity photons in cavity mode $j$, $\hat \sigma_{\xi \xi^\prime} \equiv |\xi\rangle\langle \xi^\prime|$, $\hat p$ is the center-of-mass momentum operator of the atom along the cavity axis $x$, $I_{4\times4}$ is the identity matrix in the internal atomic-state space, and H.c. stands for the Hermitian conjugate. In the rotating frame of the pumping lasers, $\tilde{\mathcal{H}} = \mathcal{U} \mathcal{H} \mathcal{U}^\dagger + i\hbar \left(\frac{\partial}{\partial t} \mathcal{U}\right)\mathcal{U}^\dagger$, using the unitary transformation 
\begin{equation}
\eqalign{\fl \mathcal{U} = \exp\Bigg\{i\bigg[\left(\frac{-\omega _{p_1}-{\omega _{p_2}}+{\omega }}2\right)\hat \sigma_{g_1g_1} +\left(\frac{-\omega _{p_1}-{\omega _{p_2}}-{\omega }}2\right)\hat \sigma_{g_2g_2}\\ +\left(\frac{\omega _{p_1}-{\omega _{p_2}}+{\omega }}2\right)\hat\sigma_{e_1e_1} +\left(\frac{-\omega _{p_1}+{\omega _{p_2}}-{\omega }}2\right)\hat \sigma_{e_2e_2} \\ + \omega_{p_1}\hat a_1^\dagger \hat a_1 + \omega_{p_2}\hat a_2^\dagger \hat a_2 \bigg]t \Bigg\},}
\end{equation}
the single-particle Hamiltonian density becomes
\begin{equation}
\eqalign{\fl
 \tilde{\mathcal{H}} = \frac{\hat p^2}{2M}I_{4\times4} +{\hbar \Delta_{g_1}} \hat \sigma_{g_1g_1}+{\hbar \Delta_{g_2}}\hat \sigma_{g_2g_2} -\hbar \Delta_{e_1} \hat \sigma_{e_1e_1} -\hbar \Delta_{e_2} \hat \sigma_{e_2e_2} - \hbar \Delta_{c_1}\hat a^\dagger_1 \hat a \\  -\hbar \Delta_{c_2} \hat a^\dagger_2 \hat a_2 + \hbar \Big[\mathcal{G}_1(x)\hat a_1 \hat \sigma_{e_1g_1} + \mathcal{G}_2(x)\hat a_2 \hat \sigma_{e_2g_2} + \mathrm{H.c.} \Big]\\ + \hbar\Big[ \Omega_1 \hat \sigma_{e_1g_1} +\Omega_2 \hat\sigma_{e_2g_2} +\mathrm{H.c.}\Big] + \hbar \Big[ \Omega \hat \sigma_{g_1g_2} + \mathrm{H.c.}\Big],}
\end{equation}
where we have defined $\Delta_{g_1} \equiv \omega_{g_1}+(\omega_{p_1}+\omega_{p_2}-\omega)/2$, $\Delta_{g_2} \equiv \omega_{g_2}+(\omega_{p_1}+\omega_{p_2}+\omega)/2$, $\Delta_{e_1} \equiv -\omega_{e_1} +(\omega_{p_1}-\omega_{p_2}+\omega)/2$, $\Delta_{e_2}\equiv -\omega_{e_2}+(\omega_{p_1}+\omega_{p_2}-\omega)/2$, and $\Delta_{c_j}\equiv \omega_{p_j}-\omega_{c_j}$ as the atomic and cavity detunings with respect to the pump lasers and oscillating magnetic field. The many-body Hamiltonian is
\begin{equation}
 \hat H = \int \hat \Psi^\dagger(x) \tilde{\mathcal M}\hat \Psi(x)\,\mathrm{d}x,
\end{equation}
where $\hat \Psi(x) = (\hat\psi_{g_1},\hat \psi_{g_2},\hat \psi_{e_1}, \hat \psi_{e_2})^\intercal$ are the bosonic field operators satisfying $\left[\hat \psi_\xi(x),\hat \psi^\dagger_\xi(x^\prime)\right] = \delta_{\xi,\xi^\prime}\delta(x-x^\prime)$, and $\tilde{\mathcal{M}}$ is the matrix form of the Hamiltonian density $\tilde{\mathcal{H}}$. Using this many-body Hamiltonian, we can write the Heisenberg equations of motion of the photonic and atomic field operators
\begin{eqnarray}\label{eq:heoms}
i\hbar \frac{\partial}{\partial t} \hat a_j  = -\hbar \Delta_{c_j}\hat a_j - i\hbar \kappa \hat a_j +\hbar \int \mathcal{G}_j(x)\hat \psi_{g_j}^\dagger \hat \psi_{e_j} \mathrm{d}x,\\
 i \hbar\frac{\partial}{\partial t}\hat \psi_{g_1}  = \left(\frac{p^2}{2M}  + \hbar \Delta_{g_1}\right)\hat \psi_{g_1} + \hbar\left(\mathcal{G}_1(x) \hat a_1^\dagger +\Omega_1\right)\hat \psi_{e_1}   + \hbar\Omega \hat \psi_{g_2}, \\
 i \hbar\frac{\partial}{\partial t}\hat \psi_{g_2}  = \left(\frac{p^2}{2M}  + \hbar \Delta_{g_2}\right)\hat \psi_{g_2}+ \hbar\left(\mathcal{G}_2(x) \hat a_2^\dagger +\Omega_2\right)\hat \psi_{e_2} + \hbar\Omega \hat \psi_{g_1}, \\
i \hbar\frac{\partial}{\partial t}\hat \psi_{e_j} = \left(\frac{p^2}{2M}  - \hbar \Delta_{e_j}\right)\hat \psi_{e_j} + \hbar\left[\mathcal{G}_j(x) \hat a_j +\Omega_j\right]\hat \psi_{g_j},
\end{eqnarray}
where $-i \kappa \hat a_j$ corresponds to the decay of the cavity mode $j$. In the limit of large atomic detunings $\Delta_{e_j}$, the atomic field operators $\{\hat\psi_{e_1},\hat \psi_{e_2}\}$ of the excited states reach quickly steady states, allowing for adiabatic elimination of their dynamics. Assuming the kinetic energies are negligible with respect to $-\Delta_{e_1}$ and $-\Delta_{e_2}$, we obtain the steady-state solutions for the atomic field operators of the excited states as
\begin{eqnarray}
  \hat \psi_{e_j} = \frac{1}{\Delta_{e_j}}\left[ \mathcal{G}_j(x) \hat a_j +\Omega_j\right]\hat \psi_{g_j}.
\end{eqnarray}
If we now insert the above steady-state solutions into the Heisenberg equations of motion~(\ref{eq:heoms}), we obtain effective equations for the photonic and atomic field operators
\begin{eqnarray}
\fl  i\hbar \frac{\partial}{\partial t} \hat a_j  = \hbar\left[- \Delta_{c_j}- i\kappa + U_j\int \hat \psi_{j}^\dagger \cos^2(k_{c_j} x)\hat \psi_{j} \mathrm{d}x \right]\hat a_j+\hbar \eta_j\int \hat \psi_{j}^\dagger \cos(k_{c_j} x)\hat \psi_{j} \mathrm{d}x,\\
\eqalign{\fl i \hbar\frac{\partial}{\partial t}\hat \psi_{1}  = \left[\frac{p^2}{2M} -\hbar \gamma B_\parallel -\hbar \omega+\hbar U_1 \cos^2(k_{c_1} x) +\hbar\eta_1 \cos(k_{c_1}x)\left(\hat a_1 + \hat a_1^\dagger \right) \right]\hat \psi_{1} \\+ \hbar\Omega \hat \psi_{2},} \\
\fl i \hbar\frac{\partial}{\partial t}\hat \psi_{2}  = \left[\frac{p^2}{2M}  + \hbar\delta+ \hbar U_2 \cos^2(k_{c_2} x) +\hbar\eta_2 \cos(k_{c_2}x)\left(\hat a_1 + \hat a_1^\dagger \right) \right]\psi_{2} + \hbar\Omega \hat \psi_{1},
\end{eqnarray}
where $\hat \psi_j \equiv \hat \psi_{g_j}$, $\Delta_j \equiv \Delta_{e_j}$, $U_j \equiv \mathcal{G}_j^2/\Delta_{j}$, $\eta_j \equiv \mathcal{G}_j\Omega_j/\Delta_{j}$, and $\delta = {\Omega^2_2}/{\Delta_{2}}- {\Omega^2_1}/{\Delta_{1}} + \omega_{g_2}$. Finally, using the mean-field approximation and replacing the photonic and atomic field operators $\hat a_j$ and $\hat \psi_{g_j}$ with their corresponding averages $\alpha_j \equiv \langle \hat a_j \rangle$ and $\hat \psi_{g_j} \equiv \psi_{g_j}$, respectively
yields the following equations of motion
\begin{eqnarray} \label{eq:eoms}
  i \frac{\partial}{\partial t} \alpha_j = \left[-\Delta_c + U_0 \langle \cos^2(k_c x)\rangle_j- i \kappa\right]\alpha_j + \eta_0\langle \cos(k_c x) \rangle_j, \\
  i \hbar \frac{\partial}{\partial t} \psi_1 =\left[\frac{p^2}{2M}+V_1(x)-\hbar(\gamma B_\parallel + \omega) \right]\psi_1 +i\frac{\Omega}{2} \psi_2, \\
  i \hbar \frac{\partial}{\partial t} \psi_2 =\left[\frac{p^2}{2M}+V_2(x)+\hbar \delta\right]\psi_2 -i\frac{\Omega}{2} \psi_1.
\end{eqnarray}

\section{Cavity fields}
\label{sec:cavity-field-ss}

\noindent The equations that govern the dynamics of cavity fields are given by (for clarity of the calculations, we have assumed the balanced condition, {i.e.}, $U_1 = U_2 \equiv U_0$, $\Delta_{c_1}=\Delta_{c_2} \equiv \Delta_c$, $\omega_{c_1}=\omega_{c_2}=\omega_{c}$, and $\eta_1 = \eta_2 \equiv \eta_0$.)
\begin{equation}\label{eq:cavityfields}
 i \frac{\partial}{\partial t} \alpha_j = \left[-\Delta_c + U_0 \langle \cos^2(k_c x)\rangle_j- i \kappa\right]\alpha_j + \eta_0\langle \cos(k_c x) \rangle_j.
\end{equation}
Assuming that we are deep in the self-ordered regime, the two integrals for spinor component $j=1$ read
\begin{eqnarray}\label{eq:int1}
 \langle \cos(k_c x) \rangle_1 = \cos^2 (\phi/2) \int \hat \psi_1^\dagger(x) \cos(k_c x) \hat \psi_1(x)\,\mathrm{d}x = - N \cos^2 (\phi/2) ,\\
 \langle \cos^2(k_c x)\rangle_1 = \cos^2 (\phi/2) \int \hat \psi_1^\dagger(x) \cos^2(k_c x) \hat \psi_1(x)\,\mathrm{d}x = N\cos^2 (\phi/2), 
\end{eqnarray}
and for $j = 2$
\begin{eqnarray}\label{eq:int2}
 \langle \cos(k_c x) \rangle_2 = \sin^2 (\phi/2) \int \hat \psi_2^\dagger(x) \cos(k_c x) \hat \psi_2(x)\,\mathrm{d}x = - N\sin^2 (\phi/2),\\
 \langle \cos^2(k_c x)\rangle_2 = \sin^2 (\phi/2) \int \hat \psi_2^\dagger(x) \cos^2(k_c x) \hat \psi_2(x)\,\mathrm{d}x = N\sin^2 (\phi/2), 
\end{eqnarray}
where $\phi$ defines the population imbalance
\begin{equation}\label{eq:relativespinor}
 \delta N = N \cos^2(\phi/2) - N \sin^2(\phi/2) = N \cos \phi.
\end{equation}
Note that because of the spontaneous symmetry breaking, the integrals $\langle \cos(k_c x) \rangle$ can also switch signs. This however does not affect the time evolution.

If we now substitute equations~(\ref{eq:int1})~and~(\ref{eq:int2}) into equation~(\ref{eq:cavityfields}), we obtain
\begin{eqnarray}
 i \frac{\partial}{\partial t} \alpha_1 = \left[-\Delta_c + N U_0 \cos^2(\phi/2)- i \kappa\right]\alpha_1 - N \eta_0\cos^2(\phi/2), \\
  i \frac{\partial}{\partial t} \alpha_2 = \left[-\Delta_c + N U_0 \sin^2(\phi/2)- i \kappa\right]\alpha_2 - N \eta_0\sin^2(\phi/2).
\end{eqnarray}
If the population imbalance is a function of time $\phi = \Omega (t-t_0) = \Omega \tilde{t}$, where $t_0$ sets the initial population imbalance, the above equations become
\begin{eqnarray}
 i \frac{\partial}{\partial t} \alpha_1 = \left[-\Delta_c + N U_0 \cos^2(\Omega \tilde{t}/2)- i \kappa\right]\alpha_1 - N \eta_0\cos^2(\Omega \tilde{t}/2), \\
  i \frac{\partial}{\partial t} \alpha_2 = \left[-\Delta_c + N U_0 \sin^2(\Omega \tilde{t}/2)- i \kappa\right]\alpha_2 - N \eta_0\sin^2(\Omega \tilde{t}/2).
\end{eqnarray}
In order to come up with an approximate solution, we assume that the dispersive shift is negligible with respect to the cavity detuning $\Delta_c - N U_0 \approx \Delta_c $, so that we get
\begin{eqnarray}
 i \frac{\partial}{\partial t} \alpha_1 \approx -( i \kappa + \Delta_c)\alpha_1 - N \eta_0\cos^2(\Omega \tilde{t}/2), \\
  i \frac{\partial}{\partial t} \alpha_2 \approx - ( i \kappa + \Delta_c) \alpha_2 - N \eta_0\sin^2(\Omega \tilde{t}/2),
\end{eqnarray}
with steady-state solutions (note that the transient solutions can be also calculated analytically, however, for the sake of simplicity, we only provide steady-state solutions) given by 
\begin{equation}
\fl \alpha_{1/2}(t) = \frac{N \eta_0\left[-(\Delta_c +i \kappa )^2 \pm i \Omega (\Delta_c +i \kappa ) \sin ( \Omega \tilde{t}) \pm (\Delta_c +i \kappa )^2 \cos (\Omega \tilde{t})+\Omega ^2\right]}{2 (\Delta_c +i \kappa )^3-2 \Omega ^2 (\Delta_c +i \kappa )},
\end{equation}
which can be used to calculate the average number of cavity photons 
\begin{equation}
\eqalign{\fl |\alpha_{1/2}|^2 = \frac{N^2 \eta_0^2 \left[\left(-\Delta_c ^2+\kappa ^2\pm\left(\kappa ^2-\Delta_c ^2\right) \cos (\Omega \tilde{t})\pm\kappa \Omega \sin (\Omega \tilde{t})+\Omega ^2\right)^2\right]}{4 \left(\Delta_c ^2+\kappa ^2\right) \left[(\Delta_c -\Omega )^2+\kappa ^2\right] \left[(\Delta_c +\Omega )^2+\kappa ^2\right]} \\ +  \frac{N^2 \eta_0^2 \left[(2 \Delta_c \kappa (\cos (\Omega \tilde{t})\pm1)+\Delta_c \Omega \sin (\Omega \tilde{t}))^2\right]}{4 \left(\Delta_c ^2+\kappa ^2\right) \left[(\Delta_c -\Omega )^2+\kappa ^2\right] \left[(\Delta_c +\Omega )^2+\kappa ^2\right]},}
\end{equation}
and finally the relative average number of intra-cavity photons
\begin{equation}
\fl	\delta n = \frac{N^2 \eta_0^2 \left[-\kappa \Omega \left(\Delta_c ^2+\kappa ^2+\Omega ^2\right) \sin (\Omega \tilde{t})-\left(\Omega ^2 \left(\kappa ^2-\Delta_c ^2\right)+\left(\Delta_c ^2+\kappa ^2\right)^2\right) \cos (\Omega \tilde{t})\right]}{\left(\Delta_c ^2+\kappa ^2\right) \left[(\Delta_c -\Omega )^2+\kappa ^2\right] \left[(\Delta_c +\Omega )^2+\kappa ^2\right]}.
\end{equation}
Let us now consider briefly three cases from the main text. Except for the \emph{magnetic resonant} case  $|\Delta_{c}|\sim\Omega$, the contribution from the $\cos (\Omega \tilde t)$ will be dominant. In the adiabatic case $|\Delta_{c}|>\Omega$, it can be easily shown that
\begin{equation}
	\delta n \approx \frac{N^2 \eta_0^2  \cos (\Omega \tilde{t})}{\Delta_c ^2+\kappa ^2},
\end{equation}
which is the steady-state solution from equation~(\ref{eq:ssnop}) multiplied by an oscillatory term. In the second limiting case $|\Delta_{c}| < \Omega$, the term multiplying the $\cos(\Omega \tilde t)$ will change the sign to negative as now $\Delta_c^4 - \Delta_c^2\Omega^2 < 0$. This change of the sign is responsible for the $\pi/\Omega$ shift between the Rabi oscillations and oscillations of $\delta n$ (see figure \ref{fig:photonosc}). Also, as the denominator grows faster with $\Omega$ than the nominator ($\delta n \sim 1/\Omega^2$), the amplitude of photon oscillations will experience a decrease with respect to the adiabatic case. Finally in the \emph{magnetic resonant} case $|\Delta_{c}|\sim\Omega$, as $\Delta_c^4 - \Delta_c^2\Omega^2 \approx 0$, the $\cos(\Omega \tilde t)$ will be negligible with respect to the $\sin (\Omega \tilde t)$ which explains the $\pi/2\Omega$ shift in the phase seen in figure \ref{fig:photonosc}. Now, as the term in the denominator $\Delta_c - \Omega$ vanishes, the oscillations of the photon number will behave as $\delta n \sim \Omega$ which explains the initial increase of the amplitude of oscillations visible in figure 2.

Note that we have not neglected the dispersive shift in the numerical simulations in the main text. However, the numerical simulations should not differ too much from the solutions above since the maximal value of dispersive shift $N U_0$ is of the order of $\sim \omega_r$ which is much smaller than other parameters $(\Delta_c,\eta_0,\kappa) = (-3300,300,300)\omega_r$. The only apparent effect that the approximate solutions are not able to predict is the long-time damping of the relative photon-number oscillations in the \emph{magnetic resonant} case.

\section*{References}
\bibliographystyle{iopart-num.bst}
\providecommand{\newblock}{}

\end{document}